# Uniformly Loaded Rectangular Thin Plates with Symmetrical Boundary Conditions

Milan Batista

Univesity of Ljubljana,

Faculty of Maritime Studies and Transport, Slovenia

milan.batista@fpp.uni-lj.si

(Jan 2010)

## Abstract

In the article the Fourier series analytical solutions of uniformly loaded rectangular thin plates with symmetrical boundary conditions are considered. For all the cases the numerical values are tabulated.

## 1 Introduction

This article is motivated by the recent work of Lim et al. ([8]) and Shuang ([16]). The authors consider the deflection of a rectangular plate under a uniform load with symmetrical homogeneous boundary conditions along each plate edge. In general each edge may be simply supported (S), clamped (C) or free (F), so there are 21 different possible combinations of boundary conditions, which are listed in Table 1, where the Reddy notation of the boundary condition is adopted in which the consecutive pair of letters indicates a boundary condition on opposite edges ([15], p. 266). Note that the cases below the diagonal in Table 1 are obtained by rotation of a plate by $90^0$. As such the problem is one of the oldest in elasticity and it is discussed in more or less detail in any textbook on plate theory ([1], [3], [5], [10], [11], [15], [17], [18], [19], [20], [21]) and even in some books on the theory of elasticity ([2],[9],[22]). Now, for solving the problem the mentioned authors introduce the symplectic method, which reduces a two dimensional plate problem to an eigenvalue problem. Lim et al. ([8]) consider the plate with two opposite edges simply supported and the other two edges arbitrarily supported,

arXiv:1001.3016v1while Shuang ([16]) solves and discuses all the 21 boundary condition cases. Here it must be noted that the method is not quite new in plate theory since it was used, for example, by Morley ([10],[13]) to solve the clamped plate problem. As these authors indicate in their works, one of the advantages of the symplectic method in comparison to other methods is that the symplectic analysis is completely rational while the traditional approach is usually a unitization of the semi-inverse method where one should guess the trail functions which exist only for very special cases of boundary conditions. This may probably be true in general, but not for the case of a rectangular symmetrically supported plate and for all the cases of a plate simply supported on opposite edges, wherein one may obtain all the solutions from the solution of a biharmonic equation obtained by the Fourier method of separation of variables ([21]).

**Table 1.** Possible combinations of boundary conditions.
Bold indicates the cases discussed by Timoshenko et al. ([18])

| Boundary condition | SS | SC | SF | CC | CF | FF |
|---|---|---|---|---|---|---|
| SS | **SSSS** | **SSSC** | **SSSF** | **SSCC** | **SSCF** | **SSFF** |
| SC |  | **SCSC** | SCSF | **SCCC** | SCCF | SCFF |
| SF |  |  | SFSF | SFCC | SFCF | SFFF |
| CC |  |  |  | **CCCC** | **CCCF** | CCFF |
| CF |  |  |  |  | CFCF | CFFF |
| FF |  |  |  |  |  | **FFFF** |

In this article the consideration is restricted to symmetrical cases of boundary conditions: SSSS, SSCC, SSFF, CCCC, CCFF and FFFF. According to the historical notes of Love ([9]), Timoshenko et al. ([18]) and Melelsho ([12]) the first SSSS plate problem was solved by Navier (1823) by using a double trigonometric series. Later Lévy (1899) provided a single trigonometric series solution of a plate which has two opposite edges simply supported. His solution may be found in all the quoted references. The clamped plate problem was solved by Koialovich (1902), Boobnoff (1914) and Hencky (1913), essentially reducing the problem to the solution of infinite systems of algebraic equations ([9],[12]). The FFFF plate was solved by Galerkin (1915) as a limit case of a plate with elasticity supported edges ([3],[4],[18]) and Nadai ([16]). A recent solution using the symplectic method was given by Lim et al. ([10]) and using the Fourier method by the present author ([1]). For the analytical solution of the CCFF plate the present author found no historical reference so apparently the only known analytical solution is given by Shuang ([16]) using the symplectic method.

There are two aims of this paper. The first is to demonstrate that by using the solution of a biharmonic equation by the Fourier method one could obtain the solutions of a stated boundary value problem in a unified rational way. The second is to provide a set of reference values of deflection, moments and shear forces in selected referenced points of the plate. There are several reasons for yet another set of reference values. First, in the literature the reference values are scattered and are sometimes given only for a very specific example. Second, probably the most quoted referenced values are those from Timoshenko ([18]), however these reference values for displacements are given up to five decimal places, for the moment up to four and for shear forces up to three. For practical design purposes this is more than needed but for comportment with the values obtained by different methods it is sometimes not enough. Also the reference values for shear forces at the edge of the plate are given only for a simply supported plate; the reference values for the FFFF plate are given only for a square plate while the reference values for the CCFF plate are not given. It must be noted that the Shuang work ([16]) covers all the cases, yet the plate aspect ratio for all the symmetrical cases is restricted from 1 to 2, and for the FFFF plate only a square plate is considered. Also in this work the accuracy of numerical calculation is not explicitly controlled; rather the author gives consecutive results with increasing numbers of series terms.

While most of the material in this work is well known it also contains a new Fourier series analytical solution of the CCFF plate and also, in contrast to Timoshenko ([18]), Lim et al. ([8]) and Shuang ([16]), in all the discussed cases the values of shear forces at the middle of a plate clamped or simply supported edges are provided and in most cases the present values cover a wider aspect ratio of the plate – mostly from 1 to 5. Before proceeding, a note on the numerical calculation strategy for summing a series is given. In all the discussed cases the calculation was performed in quad precision and the condition for termination of series summation was $|s_{n+1} - s_n| < tol \times (1 + |s_n|)$ where $s_n$ is the value after summing $n$ terms and where $tol$ was taken $10^{-9}$, $10^{-8}$ and $10^{-7}$ for calculation of deflections, moments and shear forces, respectively, in most cases. In this way the values are accurate to eight, seven and six decimal places. In all the Tables *N* indicates the number of the series term used for calculation.

## 2 General considerations

Consider a homogeneous isotropic elastic rectangular thick plate of sides $a' = 2a$ and $b' = 2b$ subject to a uniformly distributed load *q*. The Cartesian coordinate system *Oxy* is originated at the center of the plate and the plate is orientated in the way that it occupies the region $-a \leq x \leq a$ and $-b \leq y \leq b$. The governing equation of the plate is ([18], p. 82)

$$\Delta^2 w = \frac{q}{D} \qquad (1)$$

where $\Delta \equiv \frac{\partial^2}{\partial x^2} + \frac{\partial^2}{\partial y^2}$ is a two-dimensional Laplace operator, $D \equiv \frac{h^3 E}{12(1-\nu^2)}$ is bending stiffness, *E* is the elastic module, *h* is plate thickness, $\nu$ is the Poisson ratio, and *w* is the transverse deflection of the middle plane of the plate. Once *w* is known one can calculate the bending moments $M_x$, $M_y$ and twisting moment $M_{xy}$ from ([18],pp 81)

$$M_x = -D\left(\frac{\partial^2 w}{\partial x^2} + \nu \frac{\partial^2 w}{\partial y^2}\right) \qquad M_y = -D\left(\frac{\partial^2 w}{\partial y^2} + \nu \frac{\partial^2 w}{\partial x^2}\right)$$

$$M_{xy} = -(1-\nu) D \frac{\partial^2 w}{\partial x \partial y} \qquad (2)$$

and shear forces $Q_x$, $Q_y$ ([18],pp 81) and effective shear forces $V_x$, $V_y$ from ([18], p. 84)

$$Q_x = -D\frac{\partial \Delta w}{\partial x} \qquad Q_y = -D\frac{\partial \Delta w}{\partial y} \qquad (3)$$

$$V_x = -D\left[\frac{\partial^3 w}{\partial x^3} + (2-\nu)\frac{\partial^3 w}{\partial x \partial y^2}\right] \qquad V_y = -D\left[\frac{\partial^3 w}{\partial y^3} + (2-\nu)\frac{\partial^3 w}{\partial x^2 \partial y}\right] \qquad (4)$$

The equation (1) should be solved in such a way that the boundary conditions at the edge of the plate are satisfied.

For a symmetrical boundary condition the solution of governing equation (1) should be symmetrical in *x* and *y*. The symmetrical solution of (1) obtained by the Fourier method of separation of variables ([21]) may be written in the form

$$w = w_0 + \frac{q}{D}\sum_{n=0}^{\infty}(-1)^n \left( A_n \frac{\cosh \alpha_n y}{\cosh \alpha_n b} + B_n \frac{y}{b} \frac{\sinh \alpha_n y}{\cosh \alpha_n b} \right) \cos \alpha_n x \\ + \frac{q}{D}\sum_{n=0}^{\infty}(-1)^n \left( C_n \frac{\cosh \beta_n x}{\cosh \beta_n a} + D_n \frac{x}{a} \frac{\sinh \beta_n x}{\cosh \beta_n a} \right) \cos \beta_n y \qquad (5)$$

where $w_0$ is a particular solution satisfying $\Delta^2 w_0 = \frac{q}{D}$ and where

$$\alpha_n \equiv \frac{(2n+1)\pi}{2a} \qquad \beta_n \equiv \frac{(2n+1)\pi}{2b} \qquad (n=0,1,2,...) \qquad (6)$$

The particular solution $w_0$ is taken in the form of a symmetrical polynomial of the fourth order in *x* and *y*

$$w_0 = c_0 + c_1 x^2 + c_2 y^2 + c_3 x^4 + c_4 x^2 y^2 + c_5 y^4 \qquad (7)$$

This solution must satisfy the plate equation, so

$$3c_3 + c_4 + 3c_5 = \frac{q}{8D} \qquad (8)$$

The task is now to determine the unknown coefficients in (5) by boundary conditions.

Before proceeding, a few words on the determination of the particular solution are in order. It is evident that the ease of solution of a specific plate boundary value problem depends on the particular solution $w_0$. A strategy for constructing $w_0$ which simplifies the solution is the following. Observe that any of the discussed boundary conditions involve either a prescribed zero vertical displacement and/or a zero bending moment. Since according to (5) the displacement and the bending moments are functions of $\begin{Bmatrix} \cosh \alpha_n y \\ y \sinh \alpha_n y \end{Bmatrix} \cos \alpha_n x$ and $\begin{Bmatrix} \cosh \beta_n x \\ x \sinh \beta_n x \end{Bmatrix} \cos \beta_n y$, one of these function will be zero on a particular edge and only the others will remain without hyperbolic dependence on *x* or *y*. Now if $w_0$ also satisfies one or both displacement/moment boundary condition then this boundary condition becomes homogeneous and one set of unknown coefficients may be directly expressed by another and this may simplify the future treatment of the problem. In the next section the specific problems will be discussed in some detail.

## 3 Plate with two opposite edges simply supported

Consider the plate with edges $x = \pm a$ simply supported. In this case the boundary conditions are

$$w = 0 \qquad \frac{\partial^2 w}{\partial x^2} = 0 \qquad \text{at} \quad x = \pm a \tag{9}$$

These boundary conditions are satisfied by the particular solution of the form of for a simply supported beam

$$w_0 = \frac{q}{24D}\left(x^4 - 6a^2 x^2 + 5a^4\right) \tag{10}$$

By (10) the boundary conditions (9) reduce to a homogeneous system for unknown coefficients $C_n$ and $D_n$ so the solution is

$$C_n = D_n = 0 \tag{11}$$

From the remaining two boundary conditions one may calculate the coefficients $A_n$ and $B_n$.

### 3.1 Simply supported plate (SSSS)

In this case boundary conditions which should be satisfied are

$$w = 0 \qquad \frac{\partial^2 w}{\partial y^2} = 0 \qquad \text{at} \quad y = \pm b \tag{12}$$

From the boundary condition $\frac{\partial^2 w}{\partial y^2}(x, \pm b) = 0$ one obtains

$$A_n = -B_n \left( \tanh \alpha_n b + \frac{2}{\alpha_n b} \right) \tag{13}$$

and further by expanding $w(x,b) = 0$ into a Fourier series in $\cos \alpha_n x$ one finds

$$B_n = \frac{b}{a \alpha_n^4} \tag{14}$$

The coefficients of the series converge very rapidly as $O(n^{-4})$.

**Table 2a.** Deflection, bending moment and shear force factors for uniformly loaded simply supported rectangular plate with $\nu = 0.3$

| $b/a$ | center: $x = y = 0$ | | | | | |
|---|---|---|---|---|---|---|
| | $w = \alpha q a'^4 / D$ | | $M_x = \beta q a'^2$ | | $M_y = \beta_1 q a'^2$ | |
| | $\alpha$ | $N$ | $\beta$ | $N$ | $\beta_1$ | $N$ |
| 1 | 0.00406235 | 3 | 0.0478864 | 4 | 0.0478864 | 4 |
| 3/2 | 0.00772402 | 2 | 0.0811601 | 3 | 0.0498427 | 3 |
| 2 | 0.01012866 | 2 | 0.1016831 | 2 | 0.0463503 | 2 |
| 3 | 0.01223281 | 1 | 0.1188605 | 2 | 0.0406266 | 2 |
| 4 | 0.01281865 | 1 | 0.1234586 | 1 | 0.0384150 | 1 |
| 5 | 0.01297083 | 1 | 0.1246245 | 1 | 0.0377453 | 1 |
| ∞ | 0.01302083 | – | 0.1250000 | – | 0.0375000 | – |

**Table 2b.** Shear force factors for uniformly loaded simply supported rectangular plate with $\nu = 0.3$

| $b/a$ | Simply: $x = \pm a$ $y = 0$ | | | | Simply: $x = 0$ $y = \pm b$ | | | |
|---|---|---|---|---|---|---|---|---|
| | $Q_x = \gamma q a'$ | | $V_x = \delta q a'$ | | $Q_y = \gamma_1 q a'$ | | $V_y = \delta_1 q a'$ | |
| | $\gamma$ | $N$ | $\delta$ | $N$ | $\gamma_1$ | $N$ | $\delta_1$ | $N$ |
| 1 | -0.337657 | 4 | -0.420471 | 4 | -0.337657 | 870 | -0.420471 | 981 |
| 3/2 | -0.423781 | 3 | -0.485646 | 3 | -0.364010 | 862 | -0.479617 | 961 |
| 2 | -0.465030 | 2 | -0.503354 | 2 | -0.369716 | 860 | -0.495800 | 956 |
| 3 | -0.492719 | 1 | -0.504726 | 2 | -0.371162 | 860 | -0.500852 | 955 |
| 4 | -0.498486 | 1 | -0.501815 | 1 | -0.371224 | 860 | -0.501140 | 955 |
| 5 | -0.499685 | 1 | -0.500550 | 1 | -0.371227 | 860 | -0.501155 | 955 |
| ∞ | -0.500000 | – | -0.500000 | – | -0.371227 | 870 | -0.501156 | – |

Table 2a contains non-dimensional deflection and bending moments at the center of the plate and Table 2b contains non-dimensional shear forces and reduced shear forces at the middle of the plate edges for various ratios of $b/a$. Values given in Table 2a and Table 2b match those in [18] (Table 8, p. 120) for specified decimal places. The exceptions are the value of the factor for $M_y$ for $b/a = 5$ where Timoshenko gives the value for $b/a = \infty$ and the values for $Q_y$ and $V_y$ where the values match in two decimal places. Note that the limit when $b \to \infty$ gives the following values for $Q_y$ and $V_y$ at the middle of the plate edge

$$\lim_{b\to\infty} Q_y(0, \pm b) = -\frac{8G}{\pi^4} qa \qquad \lim_{b\to\infty} V_y(0, \pm b) = -\frac{4(3-\nu)G}{\pi^2} qa \qquad (15)$$

where $G \approx 0.9159656$ is Catalan's constant. These values differ slightly from those for strip reported by Timoshenko. The values given in Table 2a also match those of Lim ([8], Table 1) for all decimal places.

**3.2 Plate with two opposite edges simply supported and the other two clamped (SSCC)**

In this case the remaining boundary conditions which should be satisfied are

$$w = 0 \qquad \frac{\partial w}{\partial y} = 0 \qquad \text{at} \quad y = \pm b \qquad (16)$$

From the condition $\frac{\partial w}{\partial y}(x, \pm b) = 0$ one finds

$$A_n = -B_n \left( \coth \alpha_n b + \frac{1}{\alpha_n b} \right) \qquad (17)$$

and further by expanding $w(x, \pm b) = 0$ into a Fourier series in $\cos \alpha_n x$ one finds

$$B_n = \frac{2b}{a} \frac{\tanh \alpha_n b}{\alpha_n^4 \left( \tanh \alpha_n b + \alpha_n b \cosh^{-2} \alpha_n b \right)} \qquad (18)$$

The coefficients of the series converge very rapidly as $O(n^{-4})$.

**Table 3a.** Deflection and bending moments in uniformly loaded rectangular plate with edges $x = \pm a$ simply supported and other clamped. $\nu = 0.3$

| b/a | center: $x = y = 0$ | | | | | |
|---|---|---|---|---|---|---|
| | $w = \alpha q a'^4 / D$ | | $M_x = \beta q a'^2$ | | $M_y = \beta_1 q a'^2$ | |
| | $\alpha$ | N | $\beta$ | N | $\beta_1$ | N |
| 1 | 0.00191714 | 3 | 0.0243874 | 4 | 0.0332449 | 4 |
| 1.5 | 0.00532645 | 2 | 0.0584804 | 3 | 0.0459444 | 3 |
| 2 | 0.00844500 | 2 | 0.0868681 | 2 | 0.0473622 | 2 |
| 3 | 0.01168129 | 2 | 0.1143571 | 2 | 0.0421263 | 2 |
| 4 | 0.01266531 | 1 | 0.1222547 | 1 | 0.0389927 | 1 |
| 5 | 0.01293098 | 1 | 0.1243191 | 1 | 0.0379205 | 1 |
| a/b | $w = \alpha q b'^4 / D$ | | $M_x = \beta q b'^2$ | | $M_y = \beta_1 q b'^2$ | |
| | $\alpha$ | N | $\beta$ | N | $\beta_1$ | N |
| 1.5 | 0.00247571 | 5 | 0.0178003 | 6 | 0.0406276 | 6 |
| 2 | 0.00261080 | 6 | 0.0141717 | 8 | 0.0420629 | 8 |
| 3 | 0.00261488 | 9 | 0.0124986 | 12 | 0.0418311 | 11 |
| 4 | 0.00260519 | 12 | 0.0124751 | 15 | 0.0416780 | 15 |
| 5 | 0.00260412 | 14 | 0.0124974 | 19 | 0.0416654 | 18 |

**Table 3b.** Shear forces in uniformly loaded rectangular plate with edges $x = \pm a$ simply supported and others clamped. $\nu = 0.3$

| b/a | Simply: $x = \pm a$ $y = 0$ | | Clamped: $x = 0$ $y = \pm b$ | | | |
|---|---|---|---|---|---|---|
| | $Q_x = \gamma q a'$ | | $M_y = \beta_2 q a'^2$ | | $Q_y = \gamma_1 q a'$ | |
| | $\gamma$ | N | $\beta_2$ | N | $\gamma_1$ | N |
| 1 | -0.244401 | 4 | -0.0698374 | 125 | -0.516468 | 1156 |
| 1.5 | -0.359450 | 3 | -0.1048591 | 124 | -0.665874 | 1103 |
| 2 | -0.431664 | 2 | -0.1190840 | 123 | -0.720916 | 1085 |
| 3 | -0.485460 | 1 | -0.1246081 | 123 | -0.741092 | 1079 |
| 4 | -0.496973 | 1 | -0.1249774 | 123 | -0.742377 | 1078 |
| 5 | -0.499371 | 1 | -0.1249988 | 123 | -0.742450 | 1078 |
| a/b | $Q_x = \gamma q b'$ | | $M_y = \beta_2 q b'^2$ | | $Q_y = \gamma_1 q b'$ | |
| | $\gamma$ | N | $\beta_2$ | N | $\gamma_1$ | N |
| 1.5 | -0.239447 | 6 | -0.0821937 | 163 | -0.524336 | 1412 |
| 2 | -0.238575 | 7 | -0.0842626 | 198 | -0.512208 | 1637 |
| 3 | -0.238568 | 11 | -0.0836256 | 259 | -0.501024 | 2012 |
| 4 | -0.238569 | 14 | -0.0833502 | 314 | -0.499927 | 2325 |
| 5 | -0.238569 | 17 | -0.0833308 | 364 | -0.499977 | 2599 |

Table 3a contains non-dimensional deflection and bending moments at the center of the plate and Table 3b contains non-dimensional bending moment and shear forces at the middle of the plate edges for various ratios of $b/a$. Values for deflection and moments given in Table 3a and Table 3b match those given by Lim et al. ([9]Table 3) and Timoshenko ([18], Table 29, p. 187) for three to four decimal places. There are some minor differences between values. For example for $b/a = 1.5$ Timoshenko reports the value 0.00531 for the deflection factor while in Table 3a this factor is 0.00533.

**3.3 Plate with two opposite edges simply supported and others free (SSFF)**

The remaining boundary conditions which should be satisfied for this case are

$$M_y = 0 \qquad V_y = 0 \qquad \text{at} \quad y = \pm b \tag{19}$$

From the condition $V_y(x, \pm b) = 0$ one obtains

$$A_n = -B_n \left( \coth \alpha_n b - \frac{1+\nu}{1-\nu} \frac{1}{\alpha_n b} \right) \tag{20}$$

and further by expanding $M_y(x, \pm b) = 0$ into a Fourier series in $\cos \alpha_n x$ one finds

$$B_n = \frac{2\nu b}{(3+\nu)a} \frac{\tanh \alpha_n b}{\alpha_n^4 \left( \tanh \alpha_n b - \frac{1-\nu}{3+\nu} \alpha_n b \cosh^{-2} \alpha_n b \right)} \tag{21}$$

The coefficients of the series converge very rapidly as $O(n^{-4})$.

Table 4a contains non-dimensional deflection and bending moments at the center of the plate and Table 4b contains non-dimensional deflection, bending moment and shear forces at the middle of the plate edges for various ratios of $b/a$. Values for deflection and moments given in Table 4a and Table 4b match those given by Lim et al. ([9], Table 2) and those given by Timoshenko ([18], Table 47, p. 187) for three to four decimal places. There are some minor differences between values. For example for $b/a = 2$ at the middle of the free plate edge Timoshenko reports the value 0.01521 for the deflection factor and 0.1329 for the bending moment factor while in Table 4b these factors are 0.01520 and 0.13280

**Table4a.** Deflection and bending moments in uniformly loaded rectangular plate with edges $x = \pm a$ simply supported and other two free. $\nu = 0.3$

| | center: $x = y = 0$ | | | | | |
|---|---|---|---|---|---|---|
| $b/a$ | $w = \alpha qa'^4/D$ | | $M_x = \beta qa'^2$ | | $M_y = \beta_1 qa'^2$ | |
| | $\alpha$ | $N$ | $\beta$ | $N$ | $\beta_1 0$ | $N$ |
| 1 | 0.01309368 | 3 | 0.1225454 | 3 | 0.0270781 | 1 |
| 3/2 | 0.01289772 | 2 | 0.1228061 | 2 | 0.0338624 | 2 |
| 2 | 0.01288730 | 2 | 0.1234678 | 2 | 0.0363888 | 2 |
| 3 | 0.01295984 | 1 | 0.1244522 | 1 | 0.0374998 | 1 |
| 4 | 0.01300119 | 1 | 0.1248380 | 1 | 0.0375481 | 1 |
| 5 | 0.01301530 | 1 | 0.1249563 | 1 | 0.0375200 | 1 |
| 10 | 0.01302083 | 1 | 0.1250000 | 1 | 0.0375000 | 1 |
| $a/b$ | $w = \alpha qb'^4/D$ | | $M_x = \beta qb'^2$ | | $M_y = \beta_1 qb'^2$ | |
| | $\alpha$ | $N$ | $\beta$ | $N$ | $\beta_1 0$ | $N$ |
| 3/2 | 0.06810203 | 4 | 0.2769392 | 5 | 0.0406697 | 5 |
| 2 | 0.21940976 | 5 | 0.4945694 | 6 | 0.0485903 | 6 |
| 3 | 1.13344481 | 7 | 1.1186322 | 9 | 0.0551557 | 4 |
| 4 | 3.61447289 | 8 | 1.9933699 | 11 | 0.0569888 | 12 |
| 5 | 8.86466899 | 9 | 3.1182971 | 13 | 0.0574972 | 7 |

**Table 4b.** Deflection, bending moments and shear force in uniformly loaded rectangular plate with edges $x = \pm a$ simply supported and other two free. $\nu = 0.3$

| | Free: $x = 0$ $y = \pm b$ | | | | Simply: $x = \pm a$ $y = 0$ | |
|---|---|---|---|---|---|---|
| $b/a$ | $w = \alpha_1 qa'^4/D$ | | $M_x = \beta_2 qa'^2$ | | $Q_x = \gamma qa'$ | |
| | $\alpha_1$ | $N$ | $\beta_2$ | $N$ | $\gamma$ | $N$ |
| 1 | 0.01501126 | 9 | 0.1310877 | 45 | −0.468685 | 3 |
| 3/2 | 0.01515706 | 9 | 0.1323969 | 45 | −0.485889 | 2 |
| 2 | 0.01520217 | 9 | 0.1328020 | 45 | −0.493610 | 2 |
| 3 | 0.01521806 | 9 | 0.1329447 | 45 | −0.498676 | 1 |
| 4 | 0.01521909 | 9 | 0.1329540 | 45 | −0.499725 | 1 |
| 5 | 0.01521915 | 9 | 0.1329545 | 45 | −0.499943 | 1 |
| 10 | 0.01521916 | 9 | 0.1329545 | 45 | −0.500000 | 0 |
| $a/b$ | $w = \alpha_1 qb'^4/D$ | | $M_x = \beta_2 qb'^2$ | | $Q_x = \gamma qb'$ | |
| | $\alpha_1$ | $N$ | $\beta_2$ | $N$ | $\gamma$ | $N$ |
| 3/2 | 0.07489906 | 13 | 0.2905851 | 56 | −0.671182 | 5 |
| 2 | 0.23431397 | 15 | 0.5112501 | 65 | −0.866510 | 6 |
| 3 | 1.17335261 | 19 | 1.1378446 | 76 | −1.252217 | 8 |
| 4 | 3.69022839 | 21 | 2.0132905 | 82 | −1.636917 | 11 |
| 5 | 8.98672614 | 21 | 3.1384141 | 85 | −2.021539 | 13 |

## 4. Plate with all edges clamped (CCCC)

For a plate with clamped edges the boundary conditions are

$$w = 0 \qquad \frac{\partial w}{\partial x} = 0 \quad \text{at} \quad x = \pm a \tag{22}$$

$$w = 0 \qquad \frac{\partial w}{\partial y} = 0 \quad \text{at} \quad y = \pm b \tag{23}$$

The particular solution is constructed from the conditions $w_0(\pm a, y) = 0$ and $w_0(x, \pm b) = 0$. In this way one obtains

$$w_0 = \frac{q}{8D}(x^2 - a^2)(y^2 - b^2) \tag{24}$$

By the above particular solution the boundary conditions $w(\pm a, y) = 0$ and $w(x, \pm b) = 0$ become homogeneous and yield respectively

$$C_n = -D_n \tanh \beta_n a \qquad A_n = -B_n \tanh \alpha_n b \tag{25}$$

By expanding remaining boundary conditions $\frac{\partial w}{\partial y}(x, \pm b) = 0$ and $\frac{\partial w}{\partial x}(\pm a, y) = 0$ into a trigonometric series in $\cos \alpha_n x$ and $\cos \beta_n y$ is yielded the infinite system of algebraic equations

$$\frac{a}{b} B_n \left( \tanh \alpha_n b + \alpha_n b \cosh^{-2} \alpha_n b \right) = -\frac{4}{a} \sum_{m=0}^{\infty} D_m \frac{\alpha_n \beta_m^2}{\left( \alpha_n^2 + \beta_m^2 \right)^2} + \frac{qb}{D} \frac{1}{\alpha_n^3} \tag{26}$$

$$\frac{b}{a} D_n \left( \tanh \beta_n a + \beta_n a \cosh^{-2} \beta_n a \right) = -\frac{4}{b} \sum_{m=0}^{\infty} B_m \frac{\alpha_m^2 \beta_n}{\left( \alpha_m^2 + \beta_n^2 \right)^2} + \frac{qa}{D} \frac{1}{\beta_n^3} \tag{27}$$

The above system is equivalent to the system obtained by Hencky ([9]). The approximate solution of the system may be obtained by successive approximations. It may be shown that with the new unknowns

$$B'_n = a\alpha_n B_n / b \qquad D'_n = -b\beta_n D_n / a \tag{28}$$

the system becomes fully regular so it has a unique bounded solution which can be obtained by the method of reduction ([6],[12]).

**Table 5a.** Deflection and bending moments in uniformly loaded rectangular plate with clamped edges. $\nu = 0.3$

| b/a | center: $x = y = 0$ | | | | | |
|---|---|---|---|---|---|---|
| | $w = \alpha q a'^4 / D$ | | $M_x = \beta q a'^2$ | | $M_y = \beta_1 q a'^2$ | |
| | $\alpha$ | N | $\beta$ | N | $\beta_1$ | N |
| 1 | 0.00126532 | 3 | 0.0229051 | 4 | 0.0229051 | 4 |
| 1.5 | 0.00219652 | 5 | 0.0367714 | 6 | 0.0202680 | 6 |
| 2 | 0.00253296 | 6 | 0.0411550 | 8 | 0.0158080 | 8 |
| 3 | 0.00261723 | 9 | 0.0419013 | 11 | 0.0126928 | 12 |
| 4 | 0.00260659 | 12 | 0.0416988 | 15 | 0.0124713 | 15 |
| 5 | 0.00260423 | 14 | 0.0416666 | 18 | 0.0124941 | 19 |
| 10 | 0.00260417 | 27 | 0.0416667 | 34 | 0.0125000 | 36 |

**Table 5b.** Bending moments and shear forces in uniformly loaded rectangular plate with clamped edges. $\nu = 0.3$

| b/a | Clamped: $x = \pm a\ y = 0$ | | | | Clamped: $x = 0\ y = \pm b$ | | | |
|---|---|---|---|---|---|---|---|---|
| | $M_x = \beta_2 q a'^2$ | | $Q_x = \gamma q a'$ | | $M_y = \beta_3 q a'^2$ | | $Q_y = \gamma_1 q a'$ | |
| | $\beta_2$ | N | $\gamma$ | N | $\beta_3$ | N | $\gamma_1$ | N |
| 1 | -0.0513338 | 183 | -0.441301 | 315 | -0.0513338 | 183 | -0.441301 | 315 |
| 1.5 | -0.0756586 | 241 | -0.514332 | 427 | -0.0570242 | 177 | -0.465387 | 286 |
| 2 | -0.0828661 | 298 | -0.516015 | 566 | -0.0569865 | 17 | -0.463944 | 284 |
| 3 | -0.0837766 | 52 | -0.502594 | 844 | -0.0568857 | 176 | -0.463385 | 282 |
| 4 | -0.0833867 | 495 | -0.500035 | 1115 | -0.0568862 | 176 | -0.463390 | 280 |
| 5 | -0.0833324 | 87 | -0.499966 | -1201 | -0.0568863 | 176 | -0.463390 | 282 |
| 10 | -0.0833333 | 174 | -0.500000 | 175 | -0.0568863 | 192 | -0.463390 | 1186 |

Table 5a contains non-dimensional deflection and bending moments at the center of the plate and Table 5b contains non-dimensional bending moments and also shear forces at the middle of the plate edges for various ratios of $b/a$. The values of corresponding bending data match those given by Shuang ([16], Table 4.3) and those given by Timoshenko ([18], Table 35, p. 161) from four to five decimal places. There are some minor differences between values. For example for $b/a = 2$ Timoshenko reports the value 0.00254 for the deflection factor in the middle of the plate while in Table 5a this factor is 0.002533.

## 5. Plate with two opposite edges clamped and the other two free

Consider a plate with edges $x = \pm a$ clamped and the other edges free. The boundary conditions for this case are

$$w = 0 \quad \frac{\partial w}{\partial x} = 0 \quad \text{at} \quad x = \pm a \tag{29}$$

$$M_y = 0 \quad V_y = 0 \quad \text{at} \quad y = \pm b \tag{30}$$

The particular solution is determined by the conditions $w_0(\pm a, y) = 0$ and $M_{0y}(x, \pm b) = 0$. In this way one obtains

$$w_0 = \frac{q}{24(1-2\nu)D}(a^2 - x^2)(5a^2 - 6\nu b^2 - x^2 + 6\nu y^2) \tag{31}$$

By this the conditions $M_y(x, \pm b) = 0$ and $w(\pm a, y) = 0$ become homogeneous and yield respectively

$$A_n = -B_n\left(\tanh \alpha_n b + \frac{2}{1-\nu}\frac{1}{\alpha_n b}\right) \tag{32}$$

$$C_n = -D_n \tanh \beta_n a \tag{33}$$

By expanding the remaining boundary conditions $V_y(x, \pm b) = 0$ and $\frac{\partial w}{\partial x}(\pm a, y) = 0$ into a trigonometric series in $\cos \alpha_n x$ and $\cos \beta_n y$ is yielded the infinite system of algebraic equations

$$\frac{a}{b} B_n \left(\frac{3+\nu}{1-\nu}\tanh \alpha_n b - \alpha_n b \cosh^{-2} \alpha_n b\right) = \\ \frac{4}{(1-\nu)a}\sum_{m=0}^{\infty} D_m \frac{\beta_m^2\left[(2-\nu)\alpha_n^2 + \beta_m^2\right]}{\alpha_n(\alpha_n^2 + \beta_m^2)^2} + \frac{2(2-\nu)\nu q b}{(1-\nu)(1-2\nu)D}\frac{1}{\alpha_n^3} \tag{34}$$

$$\frac{b}{a} D_n \left( \tanh \beta_n a + \beta_n a \cosh^{-2} \beta_n a \right)$$
$$= -\frac{4}{(1-\nu)b} \sum_{m=0}^{\infty} B_m \frac{\beta_n \left[ (2-\nu)\alpha_m^2 + \beta_n^2 \right]}{\left( \alpha_m^2 + \beta_n^2 \right)^2} + \frac{2qa}{3(1-2\nu)D} \frac{a^2 \beta_n^2 - 3\nu}{\beta_n^3} \tag{35}$$

The sum of the coefficients on the right side of equation (34) is divergent so the approximate solution of the system may be obtained by solving this system by some direct method. In order to solve the system by the iterative method this equation must be modified and this may be done in the following way. Dividing the first equation by $\alpha_n$ and the second by $\beta_n$, then summing each of the results on $n$ and interchanging the order of summation on $n$ and $m$, then multiplying first by $\frac{1-\nu}{1+\nu} \frac{b}{a}$ and subtracting it from the second one finds

$$S \equiv \sum_{n=0}^{\infty} D_n \left( 1 + \nu \cosh^{-2} \beta_n a \right) = \frac{qab}{6(1-2\nu)D} \left[ \left(1-\nu+\nu^2\right)a^2 - \nu(1+\nu)b^2 \right]$$
$$-\nu \sum_{n=0}^{\infty} \frac{B_n}{\alpha_n b} \left( \frac{2}{1-\nu} \tanh \alpha_n b - \alpha_n b \cosh^{-2} \alpha_n b \right) \tag{36}$$

If one set $S$ as new the system (34) may be rewritten as

$$\frac{a}{b} B_n \left( \frac{3+\nu}{1-\nu} \tanh \alpha_n b - \alpha_n b \cosh^{-2} \alpha_n b \right) = \frac{4S}{(1-\nu)a\alpha_n}$$
$$-\frac{4}{(1-\nu)a} \sum_{m=0}^{\infty} \frac{D_m}{\alpha_n} \left[ \frac{\alpha_n^2 \left( \alpha_n^2 + \nu \beta_m^2 \right)}{\left( \alpha_n^2 + \beta_m^2 \right)^2} + \nu \cosh^{-2} \beta_m a \right] + \frac{2(2-\nu)\nu qb}{(1-\nu)(1-2\nu)D} \frac{1}{\alpha_n^3} \tag{37}$$

The obtained system of equations (34), (36) and (37) may be solved by the iterative method, however the system, as may be shown, is not regular so the existence and uniqueness of the solution is not guaranteed.

Table 6a contains nondimensional deflection and bending moments at the center of the plate and Table 6b contains nondimensional deflection, bending moment and shear forces at the middle of the plate edges for various ratios of $b/a$. The values of corresponding bending data match those given by Shuang ([16], Table 4.3) from five to six decimal places. There are some minor differences between values. For example

arXiv:1001.3016v1Shuang reports for $b/a = 1$ the value 0.00255911 for the deflection factor and the value 0.0406016 for the bending moment $M_x$ factor. In Table 6a these values are 0.00255977 and 0.0406076.

**Table 6a.** Deflection and bending moments in uniformly loaded rectangular plate with edges $x = \pm a/2$ clamped and other two free. $\nu = 0.3$

| | center: $x = y = 0$ | | | | | |
|---|---|---|---|---|---|---|
| $b/a$ | $w = \alpha q a'^4/D$ | | $M_x = \beta q a'^2$ | | $M_y = \beta_1 q a'^2$ | |
| | $\alpha$ | $N$ | $\beta$ | $N$ | $\beta_1$ | $N$ |
| 1 | 0.00255977 | 4 | 0.0406076 | 5 | 0.0109358 | 5 |
| 3/2 | 0.00257164 | 6 | 0.0411260 | 7 | 0.0123035 | 8 |
| 2 | 0.00259010 | 8 | 0.0414649 | 10 | 0.0125828 | 10 |
| 3 | 0.00260331 | 11 | 0.0416593 | 14 | 0.0125334 | 15 |
| 4 | 0.00260428 | 14 | 0.0416689 | 19 | 0.0125023 | 19 |
| 5 | 0.00260419 | 18 | 0.0416670 | 23 | 0.0124997 | 24 |
| 10 | 0.00260417 | 34 | 0.0416667 | 44 | 0.0125000 | 45 |
| $a/b$ | $w = \alpha q b'^4/D$ | | $M_x = \beta q b'^2$ | | $M_y = \beta_1 q b'^2$ | |
| | $\alpha$ | $N$ | $\beta$ | $N$ | $\beta_1$ | $N$ |
| 3/2 | 0.01309442 | 6 | 0.0906574 | 8 | 0.0195476 | 8 |
| 2 | 0.04192609 | 8 | 0.1608362 | 11 | 0.0278178 | 11 |
| 3 | 0.21632886 | 12 | 0.3627751 | 16 | 0.0414663 | 16 |
| 4 | 0.69228326 | 16 | 0.6479674 | 21 | 0.0500094 | 22 |
| 5 | 1.70513325 | 20 | 1.0167348 | 26 | 0.0544031 | 27 |
| 10 | 27.86972536 | 35 | 4.1116287 | 51 | 0.0576706 | 56 |

**Table 7b.** Deflection and bending moments in uniformly loaded rectangular plate with edges $x = \pm a/2$ clamped and other two free.

| b/a | Clamped: $x = \pm a/2 \; y = 0$ | | | | Free: $x = 0 \; y = \pm b/2$ | | | |
|---|---|---|---|---|---|---|---|---|
| | $M_x = \beta_2 q a'^2$ | | $Q_x = \gamma q a'$ | | $w = \alpha_1 q a'^4 / D$ | | $M_y = \beta_3 q a'^2$ | |
| | $\beta_2$ | N | $\gamma$ | N | $\alpha_1$ | N | $\beta_3$ | N |
| 1 | -0.08155 | 118 | -0.3462 | -1201 | 0.00290883 | 77 | 0.04342 | 170 |
| 3/2 | -0.08236 | 172 | -0.4953 | 188 | 0.00291996 | 77 | 0.04362 | 169 |
| 2 | -0.08299 | 224 | -0.4992 | 251 | 0.00291997 | 77 | 0.04362 | 168 |
| 3 | -0.08332 | 321 | -0.5001 | 379 | 0.00291979 | 77 | 0.04362 | 167 |
| 4 | -0.08334 | 410 | -0.5000 | 509 | 0.00291979 | 77 | 0.04361 | 166 |
| 5 | -0.08333 | 493 | -0.5001 | 642 | 0.00291979 | 77 | 0.04361 | 164 |
| 10 | -0.08334 | 832 | -0.5006 | -1201 | 0.00291979 | 77 | 0.04362 | 161 |
| a/b | $M_x = \beta_2 q b'^2$ | | $Q_x = \gamma q b'$ | | $w = \alpha_1 q b'^4 / D$ | | $M_y = \beta_3 q b'^2$ | |
| | $\beta_2$ | N | $\gamma$ | N | $\alpha_1$ | N | $\beta_3$ | N |
| 3/2 | -0.18469 | 81 | -0.2684 | -1201 | 0.01452547 | 110 | 0.09645 | 177 |
| 2 | -0.33449 | 60 | 0.0661 | -1201 | 0.04533656 | 132 | 0.16961 | 175 |
| 3 | -0.78129 | -1201 | 1.8399 | -1201 | 0.22657575 | 152 | 0.37671 | 167 |
| 4 | -1.42875 | -1201 | 5.7571 | -1201 | 0.71335454 | 156 | 0.66520 | 163 |
| 5 | -2.27763 | -1201 | 12.6762 | -1201 | 1.74092355 | 159 | 1.03566 | 164 |
| 10 | -9.53903 | -1201 | 121.5250 | -1201 | 28.03268847 | 191 | 4.13181 | 198 |

## 6. Plate resting on corner points with all edges free (FFFF)

The detail solution for this case may be found in ([1])

## 7. Conclusions

It was shown that the solution of the uniformly loaded rectangular plate may be obtained from the general solution of a biharmonic equation by the Fourier method of separation of variables. In the case of a plate with opposite edges simply supported one obtains well known explicate expressions for unknown coefficients of deflection series expansion while for the other cases these coefficients constitute the infinite system of algebraic equations and may be approximately calculated from the truncated system by successive approximations. For the case of the CCCC plate and the FFFF plate the successive approximations converge quickly, while for the CCFF plate the convergence is slow. Regarding the problem, there is no advantage to using the symplectic method for cases of simply supported edges since both methods lead to the same results and among them the Fourier method results are obtained directly from biharmonic equations. For other cases the symplectic method leads to a solution of the transcendental equation with complex roots and in addition to the solution of infinite system of algebraic equations for unknown eigenvalue expansion coefficients. And this is numerically no simpler than solving the infinite system of equations obtained by the Fourier method.

arXiv:1001.3016v1[11] A.E.H.Love. A Treatise on the Mathematical Theory of Elasticity. Fourth edition. Dover 1944

[12] E.H.Mansfield. The Bending and Stretching of Plates. Second Edition. Cambridge University Press 1989

[13] H.Marcus. Die Theorie elasticher Gewebe und ihre Anwendung auf die Berechnung beigsamer Platten. Springer, 1932

[14] V.V.Meleshko. Bending of an Elastic Rectangular Clamped Plate: Exact Versus Engineering Solutions. Journal of Elasticity 48, 1-50,1997

[15] L.S.D.Morley. Simple Series Solution for the Bending of a Clamped Rectangular Plate under Uniform Load. Quart. Journ. Mech. and Applied Math., 16, pp 109-113, 1963

[16] A.Nadai. Über die Beigung durchlaufender Platten und der recteckigen Platte mit freien Räanderen. ZAMM, 1925

[17] J.N.Reddy. Theory and Analysis of Elastic Plates and Shells. Second edition. CRC Press 2007

[18] C.Shuang. Symplectic Elasticity Approach for Exact Bending Solutions of Rectangular Thin Plates. Master Thesis. City University of Hong Kong, 2007

[19] R.Szilard. Theories and Applications of Plate Analysis. Wiley 2004

[20] S.Timoshenko, S.Woinowsky-Krieger. Theory of Plates and Shells. Second edition. McGraw Hill 1959

[21] A.C.Ugural. Stresses in Plates and Shells. McGraw Hill 1981

[22] E.Ventsel, T.Krauthammer. Thin Plates and Shells. Marcel Dekker 2001

[23] J.R.Vinson. Structural Mechanics: The Behavior of Plates and Shells. Wiley 1974

[24] C.Wang. Applied Elasticity. McGraw-Hill, 1953
1/27/2010  5:54:06 PM                                                                                      19